\documentclass[final,3p,times,twocolumn,numbers,sort&compress]{elsarticle}
\usepackage{graphicx}
\usepackage{amssymb}
\usepackage{hyperref}
\usepackage{amsmath}
\usepackage{threeparttable}
\usepackage{longtable}
\usepackage{booktabs}
\usepackage{url}
\usepackage{float}
\usepackage{multirow}
\begin{document}

\begin{frontmatter}


\title{An Improved Reversible Data Hiding Scheme by Changing Modification Direction of Partial Coefficients in JPEG Images}

%\title{sadffff}
\author[mymainaddress]{Yi Chen}
%\cortext[mycorrespondingauthor]{}
%\ead{yichen.research@gmail.com}
%\ead{yichen.research@gmail.com}

%\address{Address\fnref{label3}}

%\title{}

%%\address[mymainaddress]{Southwest Jiaotong University, Chengdu 611756, China}

\author[mymainaddress]{Hongxia Wang\corref{mycorrespondingauthor}}
\cortext[mycorrespondingauthor]{Corresponding author}
\ead{hxwang@swjtu.edu.cn}
%\author[mysecondaryaddress]{Hanzhou Wu}
%\author[mymainaddress]{Yanli Chen}
%\author[mymainaddress]{Yong Liu}
\address[mymainaddress]{School of Information Science and Technology, Southwest Jiaotong University, Chengdu 611756, China}
%\address[mysecondaryaddress]{Institute of Automation, Chinese Academy of Sciences (CAS), Beijing 100190, P. R. China}

\begin{abstract}
This paper first reviews the reversible data hiding scheme, of Liu et al. in 2018, for JPEG images. After that, an improved reversible data hiding scheme, in which  modification directions of partial nonzero quantized alternating current (AC) coefficients are utilized to decrease distortion  and file size increase caused by data hiding, is proposed. Experimental results have shown that the proposed scheme has indeed advantages in visual quality and  smaller increase in file size of marked JPEG images while compared to the state-of-the-art scheme with the same embedding payload so far.
\end{abstract}


\end{frontmatter}

% \linenumbers

\section{Introduction}
Reversible data hiding (RDH), also called as lossless data hiding, can embed information in the host, such as images, the marked host can be restored to the original host (``clean'') after the embedded information is extracted out. Therefore, it plays a significantly important role in medical and military fields because of its reversibility.

Up  to now, many RDH schemes have been proposed and they are mainly based on histogram shifting (HS)  \cite{2006NiRDH}, difference expansion (DE) \cite{2003TianRDE} and lossless compression \cite{2002FridrichLosslessDE}. In fact, there are a fewer researchers who pay attention to lossless compression-based RDH schemes since larger embedding payloads cannot be obtained and more significant degradation in visual quality of marked images may be caused by the mean of lossless compression at present, but we can know most researches are mainly based on HS and DE according to the literature \cite{2016ShiRDH}. Since the technologies of HS and DE are respectively proposed by Ni et al. and Tian, many improvements of them have been proposed and designed for images with different formats. However, most of them are designed for uncompressed images and many significantly successful achievements in terms of visual quality and embedding capacity of marked uncompressed images have been made in the past two decades. Hence, designing RDH schemes with good visual quality and embedding capacity for marked compressed images, e.g., JPEG images, has attracted increasingly interest from more and more researchers.

File size, visual quality and embedding capacity that are three widely used standards of evaluation must be considered when designing RDH for JPEG images. In general, the requirement of visual quality for encrypted  JPEG images is unnecessary and thus it may be easier to design a RDH algorithm for them compared with unencrypted JPEG images \cite{2018QianNewFofRDHinEjpeg}. In other words, designing RDH algorithm for unencrypted JPEG images is more challenge.

\begin{figure}[t]
\setlength{\abovecaptionskip}{0.cm}
\setlength{\abovecaptionskip}{-0.cm}
\begin{center}
\includegraphics[width=80mm]{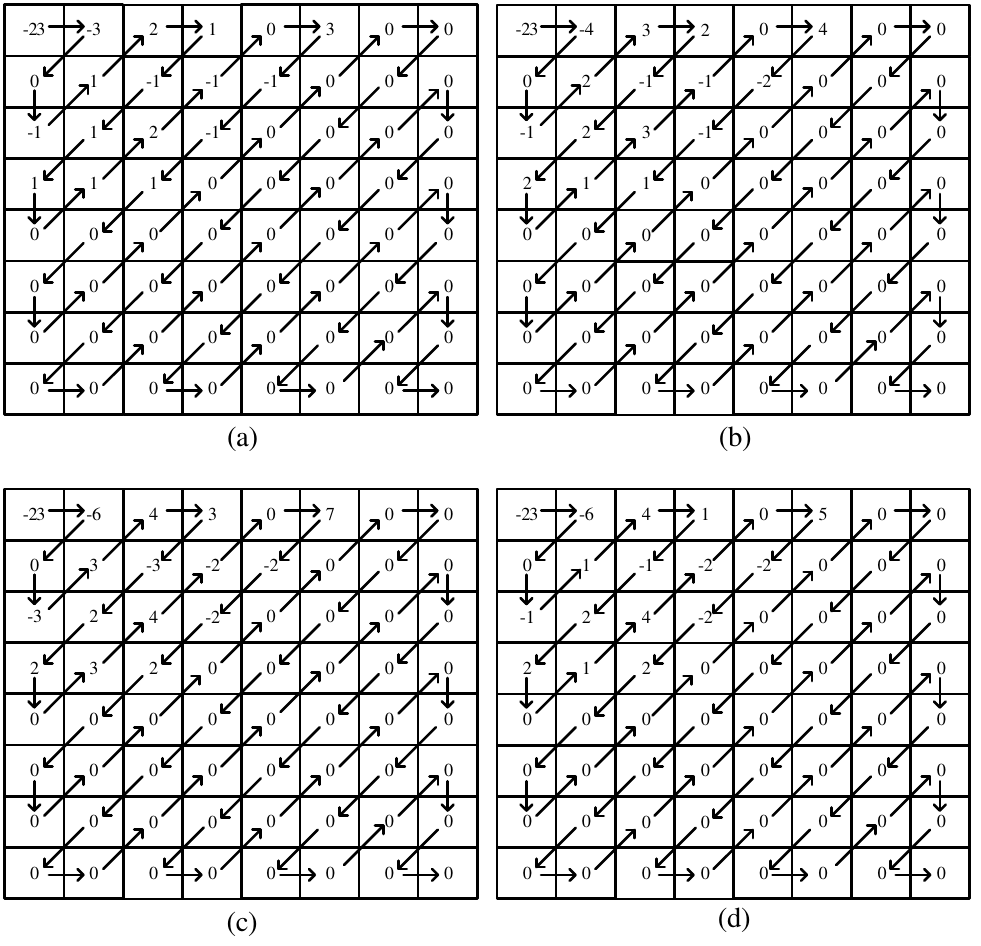}
\end{center}
\caption{Coefficient changes before and after data hiding. (a) Original JPEG quantized coefficients. (b) Marked coefficients by \cite{2016HuangRDHinJPEG} and embedded information bits: ``0110 1101 100''. (c) Marked coefficients by \cite{2018LiuRDHforJPEG} and embedded information bits: ``0110 1100 1001 000''. (d) Marked coefficients using the proposed scheme and embedded information bits: ``0110 1100 1001 000''.}
\label{ACCoefficients}
%\vspace*{-3pt}
%{\hfill\footnotesize Note how the caption is centered in the column.\hfill}
\end{figure}

Recently, Nikolaidis modified zero quantized alternating current (AC) coefficient combined with a mapping rule to propose a RDH \cite{2015NikolaidisRDHinJPEG}. However, increasing the number of nonzero quantized coefficients means that file size may significantly become larger with the increase of embedding capacity. Alternatively, Huang et al. proposed an excellent RDH scheme based on HS for JPEG in \cite{2016HuangRDHinJPEG}. In this scheme, only the AC coefficients with magnitude 1 are exploited to hide secret information bits and other nonzero AC coefficients are shifted to vacate room for hiding data. In addition, they only take advantage of the blocks with more zero coefficients because it may lead to less invalid shifting and thus obtain higher visual quality of marked JPEG images. For instance, Fig. \ref{ACCoefficients}(a) denotes a selected block with quantized coefficients for data hiding and it becomes Fig. \ref{ACCoefficients}(b) with the scheme of Huang et al. \cite{2016HuangRDHinJPEG}. Furthermore, Liu et al. utilized all nonzero AC coefficient to propose a simple RDH
scheme with higher embedding capacity for JPEG images. Besides, the scheme obtained smaller increase in file size of marked JPEG images compared with that of Huang et al.'s scheme when using special encoding principle, which is addressed in detail in \cite{2018LiuRDHforJPEG}. With their scheme, all AC quantized coefficients are changed from Fig. \ref{ACCoefficients}(a) to Fig. \ref{ACCoefficients}(c). To our best knowledge, Liu et al.'s scheme is the state-of-the-art in the embedding capacity at present. However, we are inspired by Liu et al.'s scheme to propose a better RDH scheme with the same embedding capacity but better visual quality and smaller increase in file size of marked JPEG images in this paper.
\section{The proposed method}\label{Proposed}
The proposed scheme is very simple and the difference between it and \cite{2018LiuRDHforJPEG} is that we make full use of another modification direction, which is different from that of \cite{2018LiuRDHforJPEG}, to decrease the increase of file size and keep better visual quality of marked JPEG image. More details are demonstrated in the following.
\subsection{Data Embedding}
\begin{figure}[!thb]
\setlength{\abovecaptionskip}{0.cm}
\setlength{\abovecaptionskip}{-0.cm}
\begin{center}
\includegraphics[width=80mm]{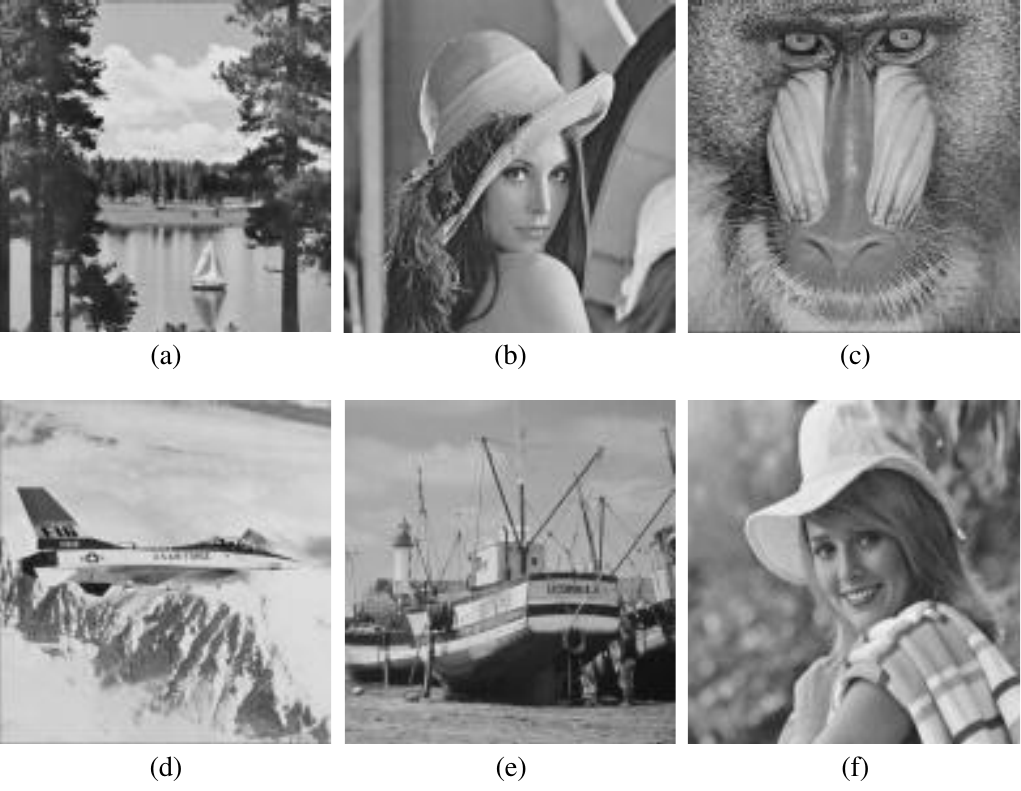}
\end{center}
\caption{Test images. (a) Lake. (b) Lena. (c) Mandrill. (d) Jetplane. (e) Boat. (f) Elaine.}
\label{JPEGOr}
%\vspace*{-3pt}
%{\hfill\footnotesize Note how the caption is centered in the column.\hfill}
\end{figure}
JPEG images is firstly entropy decoded, divided into $8 \times 8$ non-overlapping blocks and obtained quantized coefficients by the sender. After that, all nonzero quantized AC coefficients are utilized to embed data. The algorithm of the proposed scheme is stated as follows.
\begin{equation}
\bar{C} =
\begin{cases}
2 \times C \;\;\textrm{if} \;\;S = 0\\
2 \times C -sign(C)\;\;\textrm{if}\;\;S = 1\\
\end{cases}
\end{equation}
where $C$  and $S$ denote a nonzero quantized AC coefficient and  a to-be-embedded bit, respectively, and
\begin{equation}
sign(x) =
\begin{cases}
1 \;\; \textrm{if}\;\; x > 0\\
-1\;\; \textrm{if}\;\; x < 0\\
\end{cases}
\end{equation}

When our proposed scheme is implemented in an embeddable block, such as Fig. \ref{ACCoefficients}(a), these coefficients will be changed corresponding to Fig. \ref{ACCoefficients}(d).
\subsection{Data Extraction and JPEG Images Recovery}
\begin{table*}[t]
  \caption{Comparisons of embedding capacity (bits)}\label{EC}
  \centering
  \begin{tabular}{lccccccccccc}
  \hline
  \multirow{2}{1.5cm}{Images}&\multicolumn{3}{c}{QF=50}&\multicolumn{1}{c}{}&\multicolumn{3}{c}{QF=70}&\multicolumn{1}{c}{}&\multicolumn{3}{c}{QF=90}\\
   \cline{2-4}\cline{6-8}\cline{10-12}
                       &\cite{2016HuangRDHinJPEG}&\cite{2018LiuRDHforJPEG}&Proposed&&\cite{2016HuangRDHinJPEG}&\cite{2018LiuRDHforJPEG}&Proposed&&\cite{2016HuangRDHinJPEG}&\cite{2018LiuRDHforJPEG}&Proposed\\

  \hline
  Lake     & 20564  &37340  &37340  &&26814  &51901  &51901   &&49799  &101636  &101636\\
  Lena     & 14425  &24689  &24689  &&20250  &36056  &36056   &&38649  &74694   &74694 \\
  Mandrill & 35047  &64116  &64116  &&43070  &86722  &86722   &&62857  &149573  &149573\\
  Jetplane & 14880  &26989  &26989  &&19734  &37733  &37733   &&33653  &71506   &71506 \\
  Boat     & 17063  &30548  &30548  &&21697  &42100  &42100   &&36337  &78577   &78577 \\
  Elaine   & 17196  &26250  &26250  &&27129  &42071  &42071   &&58531  &100400  &100400\\
  \hline
\end{tabular}
\end{table*}
\begin{figure*}[htbp]
\setlength{\abovecaptionskip}{0.cm}
\setlength{\abovecaptionskip}{-0.cm}
\begin{center}
\includegraphics[width=150mm]{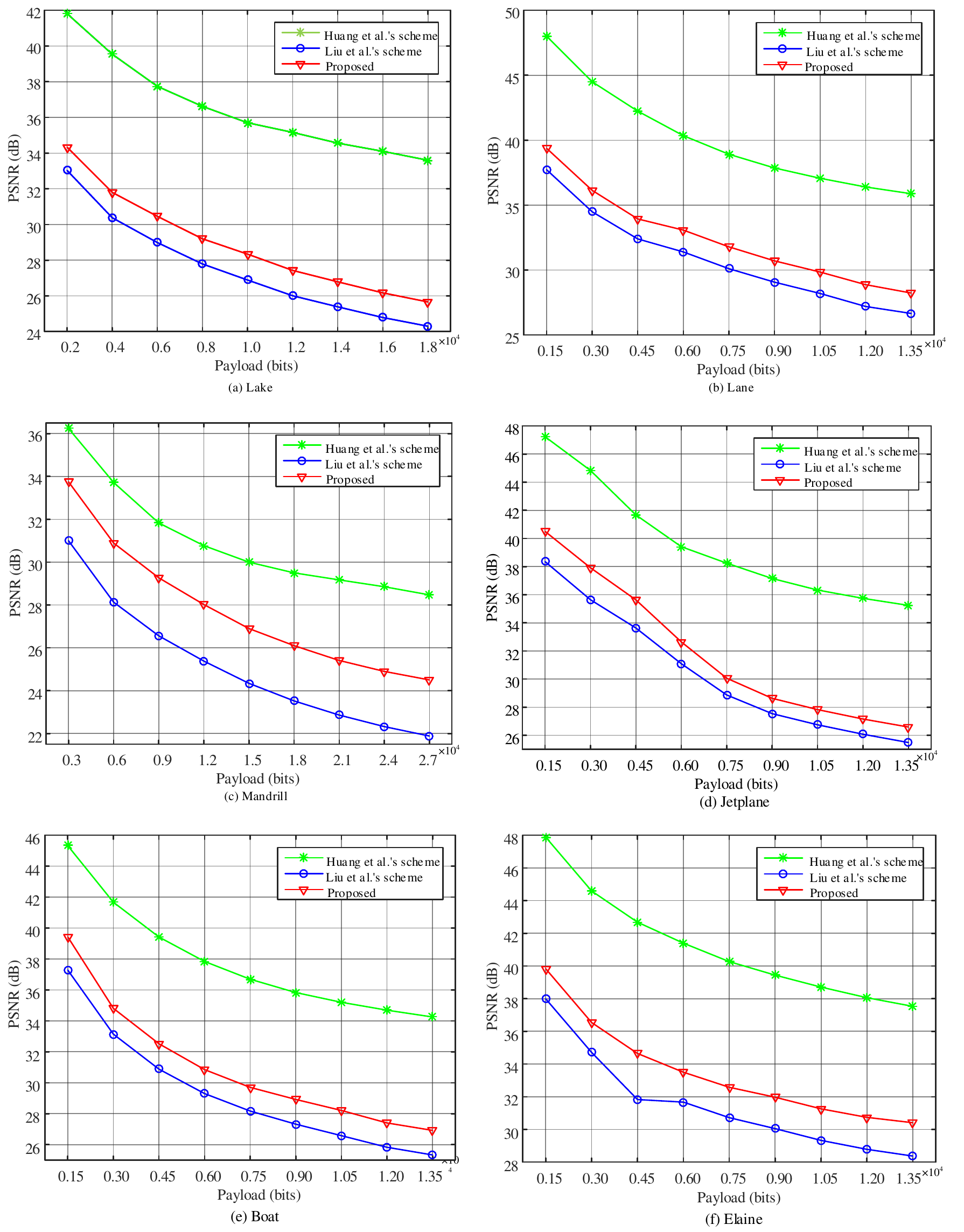}
\end{center}
\caption{Comparison of PSNR variations of marked JPEG images (QF = 50).}
\label{ComparisonOfPSNR}
%\vspace*{-3pt}
%{\hfill\footnotesize Note how the caption is centered in the column.\hfill}
\end{figure*}
\begin{figure*}[htbp]
\setlength{\abovecaptionskip}{0.cm}
\setlength{\abovecaptionskip}{-0.cm}
\begin{center}
\includegraphics[width=150mm]{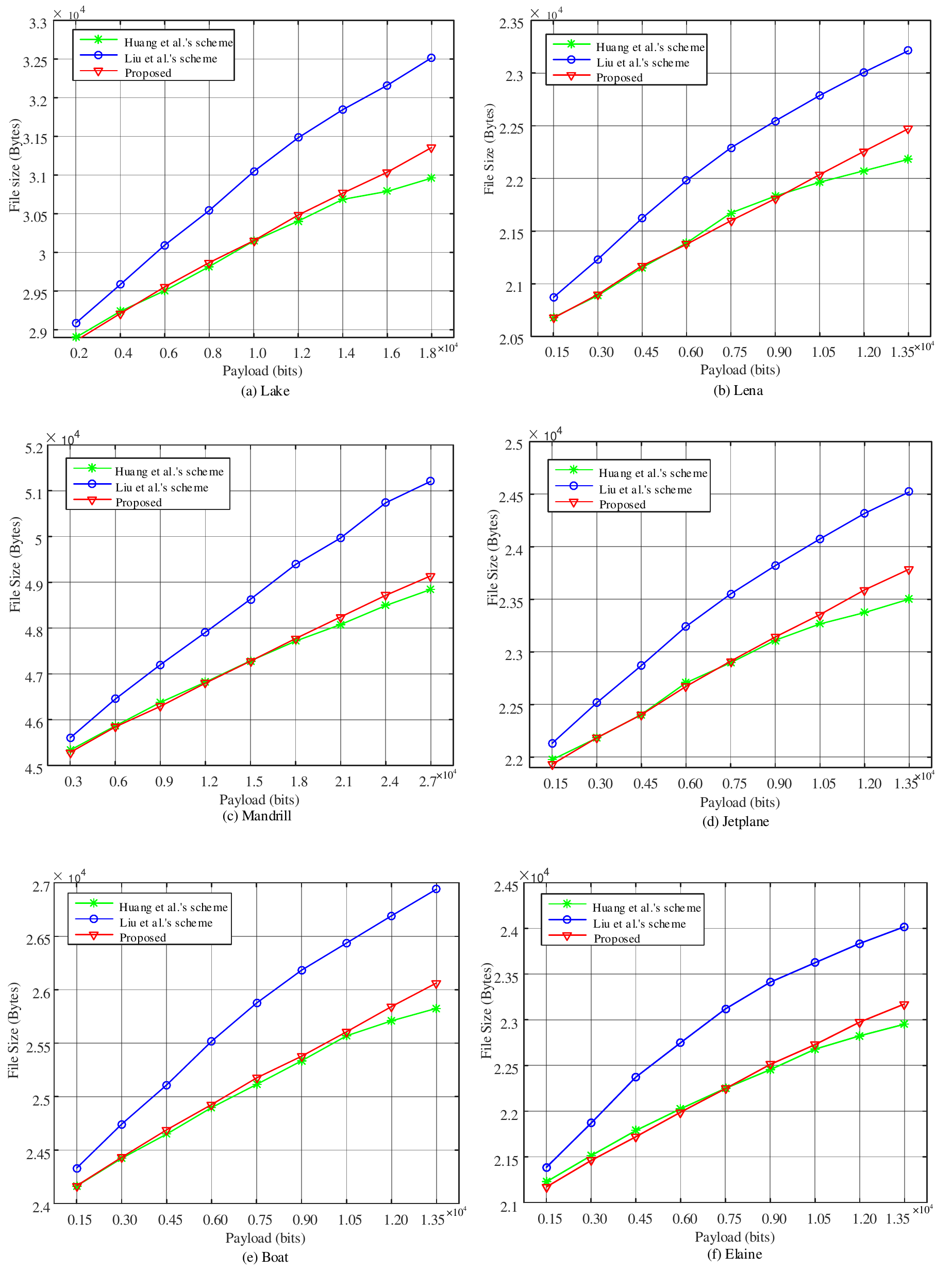}
\end{center}
\caption{Comparison of file size of marked JPEG images (QF = 50).}
\label{FileSize}
%\vspace*{-3pt}
%{\hfill\footnotesize Note how the caption is centered in the column.\hfill}
\end{figure*}

The recipient decodes the marked JPEG images after he/she receives them and obtains nonzero quantized AC coefficients. The embedded information bit is extracted from a nonzero coefficient $\bar{C}$ and the coefficient of JPEG images are restored in the following.
\begin{equation}
S^\prime =
\begin{cases}
0\;\;\textrm{if}\;\;\bar{C}\;\;\textrm{is even.}\\
1\;\;\textrm{if}\;\;\bar{C}\;\;\textrm{is odd.}\\
\end{cases}
\end{equation}

\begin{equation}
C^\prime =
\begin{cases}
\bar{C}/2\;\;\textrm{if}\;\;\bar{C}\;\;\textrm{is even.}\\
(\bar{C} + sign(C))/2\;\;\textrm{if}\;\;\bar{C}\;\;\textrm{is odd.}\\
\end{cases}
\end{equation}
where $S^\prime$ and $C^\prime$ denote an extracted message bit and a restored AC coefficient, respectively.
\section{Experimental results and analysis}\label{Results}

With the function of ``imwrite'' of ``MATLAB'', six standard 512 $\times$ 512 grayscale images, including ``Lake'', ``Lena'', ``Mandrill'', ``Jetplane'', ``Boat'' and ``Elaine'', are converted to JPEG images with different quality factors, i.e., QF = 50, 70 and 90, shown in Fig. \ref{JPEGOr}, which are used in experiments to evaluate the performance of our proposed scheme.

We make use of Table \ref{EC} to demonstrate the embedding capacities of Huang et al.'s scheme, Liu et al.'s scheme and our proposed scheme on these test images. Clearly, Liu et al.'s scheme and our proposed scheme can obtain higher embedding capacity when compared with Huang et al.'s scheme. Liu et al.'s scheme and our proposed scheme make full use of each nonzero AC coefficient to carry secret bits. However, in Huang et al.'s scheme only the nonzero AC coefficients with magnitude 1 are exploited to carry secret bits and other nonzero AC coefficients are shifted to vacate room for reversibility. Therefore, the first two schemes have indeed an advantage in the embedding capacity when compared with Huang et al.'s scheme and their embedding capacities are approximately twice as much as that of Huang et al.'s scheme.

Figs. \ref{ComparisonOfPSNR}-\ref{FileSize} are exploited to discuss the visual quality and the increased file size of marked JPEG images. We just give the experimental results on these test images with QF = 50 in this paper and we think they can represent other cases, i.e., test images with other different quality factors. Obviously, Liu et al.'s scheme and our proposed scheme degradate more significantly compared with Huang et al.'s scheme according to Fig. \ref{ComparisonOfPSNR}. This is because nonzero AC coefficients are at most increased or decreased by 1 in Huang et al.'s scheme but not in our proposed scheme and Liu et al.'s scheme. Furthermore, the visual quality of our proposed scheme is improved when compared with that of Liu et al.'s scheme. Alternatively, we give Fig.\ref{FileSize} to compare the increased file size of marked JPEG images using the three schemes under the same embedding capacity. Obviously, using Liu et al.'s scheme leads to significant increased file size of marked JPEG images when compared with our proposed scheme and Huang et al.'s scheme. From Fig. \ref{FileSize}(a-f), we can observe that using our proposed scheme results in the increased file size close to that using Huang et al.'s scheme. When there exist a lot of nonzero AC coefficients with magnitude of $\le$ 2, our proposed scheme may have an advantage. For instance, if a nonzero AC coefficient is with the value of 2. By Huang et al.'s scheme, it must become 3. Moreover, it will become 4 or 5 which corresponds to the to-be-embedded bit is ``0'' or ``1'' with Liu et al.'s scheme. In contrast, it will become 4 or 3 corresponding to the to-be-embedded bit ``0'' or ``1'' by our proposed scheme. That is to say, our proposed scheme can embed one secret bit sometimes the modification is identical but Huang et al.'s scheme cannot. In result, our proposed scheme leads to the less number of modified coefficients compared with Huang et al.'s scheme when the embedding capacity is same, which makes the increased file sizes very close. Based on the above-mentioned analysis, we can know our proposed scheme outperforms Liu et al.'s scheme.
\section{Conclusions}\label{Conclusions}
We propose a novel and simple reversible data hiding scheme for JPEG images in this paper. Compared with the state-of-the-art scheme, our proposed scheme can (1) obtain the same embedding capacity. In other words, each nonzero quantized AC coefficient can carry one information bit; (2) improve the visual quality of marked JPEG images; (3) and decrease the increased file size of marked JPEG images. In addition, our proposed scheme can keep the increased file size very close to that of the state-of-the-art HS-based scheme while embedding the same embedding payloads. In fact, the increased file size of JPEG images using our proposed scheme is less than that of Huang et al.'s scheme when the embedding payloads is smaller, e.g., $<$ 0.60 $\times 10^{4}$ bits like Fig. \ref{FileSize}(f).
\section*{Acknowledgment}

This work was supported by the National Natural Science Foundation of China (NSFC) under the grant No. U1536110.


%% References


\bibliographystyle{elsarticle-num}
\bibliography{ReferenceChen}






% \end{thebibliography}

\end{document}